\documentclass[aps]{revtex4}

\include{psfig}  
\begin{document}
\title{\bf Surface impedance of superconductive thin films as a 
function of frequency in microwave range}
\author{S.Sarti}
\affiliation{Dipartimento di Fisica and Unit\`{a} INFM, Universit\'a di Roma {\it La Sapienza},\\
 P.le Aldo Moro 2, I-00185, Roma, Italy}
\author{C.Amabile}
\affiliation{Dipartimento di Fisica and Unit\`{a} INFM, Universit\'a di Roma {\it La Sapienza},\\
 P.le Aldo Moro 2, I-00185, Roma, Italy}
\author{N.Tosoratti}
\altaffiliation{Present adress: Istituto per i processi chimico-fisici, area di ricerca CNR di Pisa, V. G. Moruzzi 1, 56124 Pisa} 
\affiliation{Dipartimento di Fisica and Unit\`{a} INFM, Universit\'a di Roma {\it La Sapienza},\\
 P.le Aldo Moro 2, I-00185, Roma, Italy}
\author{E.Silva}
\affiliation{Dipartimento di Fisica ``E.Amaldi'' and Unit\`{a}
INFM,\\ Universit\`{a} ``Roma Tre'', Via della Vasca Navale 84, 00146
Roma,Italy}

\date{\today} 

\begin{abstract}
We report measurements of the complex resistivity in $YBCO$ and $MgB_2$ thin films over a continuous frequency spectrum in the microwave range, making use of a Corbino disk geometry. The paper mainly focuses on the extraction of the resistivity from raw data, displaying data analisys procedure and its limits of validity. We obtain and show resistivity curves as a function of frequency and temperature denoting a frequency dependent widening of the superconductive transition.
\end{abstract}

\maketitle
\section{Introduction}

The microwave response of superconductors is of great interest in several  temperature and magnetic field regions, since it carries information about both macroscopic and microscopic quantities, such as resistivity or electromagnetic penetration depth.

Most of available experimental data are restricted to fixed frequencies (resonant cavity techniques), whereas a study over a continuous frequency spectrum would allow for a deeper investigation of the dynamical behaviour of the system. An 
experimental technique which makes it possible to evaluate the complex 
surface impedance $Z_{s}=R_{s}+iX_{s}$ of a superconducting film as a function of frequency has been proposed in the past by some groups \cite{anlage}\cite{toso1}. The technique is based on the measurement of the reflection coefficients at the input of a coaxial line terminated by the film (Corbino disk). 

Extracting the contribution of the sample from the measured data is a critical step during data analisys: in fact, the contribution of the coaxial line is in general of the same order of magnitude of that of the sample.
The main purpose of this paper is to present the most recent results on this experimental procedure, as obtained in our laboratory.

\section{Experimental setup}
The electromagnetic field is generated and analized by a Vector Network Analizer (VNA), connected to the sample by a commercial coaxial cable (cutoff frequency: 60 GHz), about 3.5 meters long (1 meter inside the cryostat and 2.5 meters outside). Such a large distance is needed since the experimental setup has been studied for high magnetic field measurements and the VNA needs to be placed as far as possible from the magnet. On the other hand such a high length results in a heavy attenuation of the signal, especially at higher frequencies, which prevented us to perform measurements above 35 GHz and resulted in highly noisy measurements above 20 GHz.

The sample surface impedance is probed as a function of frequency through a Corbino disk geometry, in which the sample shortcircuits the above mentioned cable.
Temperature is varied employing a helium flow cryostat. The connection between the cable and the sample is realized through a double spring method, extensively described in \cite{toso1}.  With respect to previous realizations, a thin indium ring has been placed between the sample and the outer connector of the cable, to reduce the possible damage of the sample. However this was not sufficient to suppress a detectable contact capacitance. The capacitance effect results to be smaller for the $MgB_2$ film than for the $YBCO$ film, suggesting that the sample surface plays a relevant role in the overall contact realization. As a result of this spurious capacitance, data on $YBCO$ are not reliable below 6 GHz, while in 
the case of $MgB_2$ the capacitance masks data only below 0.5 GHz, so that the actual frequency range over which measurements will be presented in the following of the paper is 
limited to $6\div 20$ GHz for $YBCO$ and $0.5 \div 20$ GHz for $MgB_2$. 

The samples here considered are a thin (2200\AA), square (l = 10 mm) $YBCO$ film,
grown by planar high oxygen pressure d.c, sputtering on a $LaAlO_{3}$ substrate \cite{crescitaYBCO}, and a thin (1500\AA), square (l= 5 mm) $MgB_2$ film, grown by pulsed laser deposition with a conventional two step technique on a sapphire substrate \cite{crescitaMgB2}. 

\section{Measurement technique}

\subsection{Equations and approximations}

As discussed in the introduction, the problem issued in this paper is how to extract the surface impedance $Z_s$ of the sample  from the reflection coefficients $\Gamma_m(\nu,T)$ measured at the beginning of the coaxial cable. $Z_s$ is defined as
the ratio between transverse electric and magnetic field components at the sample surface. It is possible to demonstrate \cite{toso1},\cite{anlage} that $Z_s(\nu,T)$ is related to the reflection coefficient at the sample surface $\Gamma_0(\nu,T)$ through the relation
\begin{equation}
\label{Zs}
Z_s(\nu,T) =  Z_0\frac{1+\Gamma_0(\nu,T)}{1-\Gamma_0(\nu,T)}
\end{equation}
where $Z_0$ is the characteristic impedance of the dielectric filling the cable. We estimate it in $377\Omega$ (vacuum impedance) being the last section of the cable is not filled with the dielectric. Due to the very small thickness of the samples, the thin film approximation is always valid \cite{silva} and one has $Z_{s} \simeq \rho/d$, where $\rho$ is the complex resistivity of the sample.

In principle, the values of $\Gamma_0(\nu,T)$ can be extracted from the measured reflection 
coefficient at the VNA input $\Gamma_m(\nu,T)$ through the standard relation \cite{Collin}
\begin{equation}
\label{gammam}
\Gamma_m = E_d+\frac{E_r\Gamma_0}{1-E_s\Gamma_0}
\end{equation}
where $E_r$, $E_s$ and $E_d$ are complex parameters taking into account cable effects. They depend on frequency and on the temperature of the whole cable. These three cable coefficients can be determined by measuring the response of three known loads replacing the sample (line calibration) but in our case the experimental setup doesn't allow for a full calibration of the cable during measuring sessions since the measurement cell is inaccessible and calibration standards are not available at the temperatures of interest in this work. 
However, we performed daily calibration of the section of the cable external to the cryostat, hence from here on we will no further consider it, as if the VNA were directly connected to the internal, one-meter long, cable. This means that the cable coefficients and reflection coefficients dealt with in the following are only relative to the internal cable.

\begin{figure}
\centerline{\psfig{figure=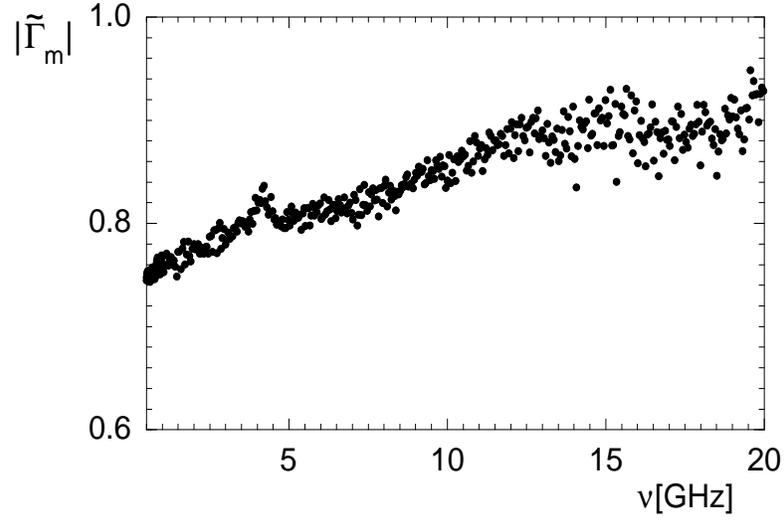,height=7.cm,width=10.5cm,clip=,angle=0.}}
\caption{Typical curve of $|\widetilde{\Gamma}_{m}|(\nu,T)$ at fixed temperature T=40K as a function of frequency for $MgB_{2}$. Since $|\Gamma_{0}|$ is reasonably constant with frequency at this temperature we attribute the frequency dependence of $|\widetilde{\Gamma}_{m}|(\nu,40K)$ to a frequency dependence in $|\alpha(\nu,40K)|$.}
\label{Gmtilde(nu)}
\end{figure}

Before the beginning of the measuring session we measure and store the coefficients of the internal cable at room temperature $T_{r}$. We then define $\widetilde{\Gamma}_{m}(\nu,T)$ as:
\begin{equation}
\label{gammamtilde}
\widetilde{\Gamma}_m(\nu,T)=\frac{\Gamma_m(\nu,T)- E_{d}(\nu,T_{r})}{E_{r}(\nu,T_{r}) +E_{s}(\nu,T_{r})\left(\Gamma_m(\nu,T)-E_{d}(\nu,T_{r})\right)}
\end{equation}
As can be seen inverting equation \ref{gammam} at $T=T_{r}$ we have $\widetilde{\Gamma}_{m}(\nu,T_{r})=\Gamma_0(\nu,T_{r})$. To find a relation between $\widetilde{\Gamma}_{m}(\nu,T)$ and $\Gamma_0(\nu,T)$ at $T<T_{r}$, we put $\Gamma_{m}(\nu,T)$ as obtained from equation \ref{gammam} into equation \ref{gammamtilde} (we will omit in the following the explicit frequency dependence for the sake of compactness):
\begin{equation}
\label{gamma0prima}
\Gamma_{0}=\frac{E_{d}(T_{r})-E_{d}(T)+\frac{E_{r}(T_{r})\widetilde{\Gamma}_{m}}{1-E_{s}(T_{r})\widetilde{\Gamma}_{m}}}
{E_{r}(T)+E_{s}(T)\left[E_{d}(T_{r})-E_{d}(T)+\frac{E_{r}(T_{r})\widetilde{\Gamma}_{m}}{1-E_{s}(T_{r})\widetilde{\Gamma}_{m}}\right]}
\end{equation}
We expect cable coefficients not to change by more than 10\% for any variation of the temperature profile in the cable. Moreover, during the room temperature calibrations, we found $|E_{d}|$ and $|E_{s}|$ to be quite generally no more than 10\% of $|E_{r}|$, which is instead of the same order of $|\widetilde{\Gamma}_{m}|\simeq 1$. This means that $|E_{d}(T_{r})-E_{d}(T)|\simeq 10^{-2}|E_{r}\widetilde{\Gamma}_{m}|$ and we can neglect it in equation \ref{gamma0prima}, obtaining after some algebra
\begin{equation}
\label{gamma0seconda}
\Gamma_{0}\simeq\frac{\frac{E_{r}(T_{r})}{E_{r}(T)}\widetilde{\Gamma}_{m}}
{1+\widetilde{\Gamma}_{m}{E_{s}(T)}\left[\Delta E_s-\Delta E_r\right]}
\end{equation}
Where $\Delta E_{s,r} = (E_{s,r}(T)-E_{s,r}(T_r))/E_{s,r}(T)$. Again $E_{s}\widetilde{\Gamma}_{m}\simeq 0.1$ and $\Delta E_{s,r}\simeq 0.1$ so that one gets, to within a few percent of precision,
\begin{equation}
\label{gamma0finale}
\Gamma_{0}\simeq\frac{E_{r}(T_{r})}{E_{r}(T)}\widetilde{\Gamma}_{m}\doteq\frac{\widetilde{\Gamma}_{m}}{\alpha(T)}
\end{equation}
We want here to stress that this just defined coefficient $\alpha(\nu,T)$ depends in a non-trivial way upon the whole temperature profile of the cable so that no practical measurement of it is possible. Nevertheless, when performing measurements at low T $(T\lesssim 100K)$, it is reasonable to expect that it is almost constant when varying the temperature of the sample of a few tens of Kelvin; the cable is in fact thermally constrained at room temperature at the entrance of the cryostat so that the temperature profile of the cable only slightly changes for such variations of the temperature of the sample.
We can also roughly estimate the frequency dependence of $\alpha(\nu,T)$ considering that it is the ratio between two transmission coefficients at different temperatures. Since $E_{r}(\nu,T)$ takes into account the attenuation of the field due to losses through the dielectric and on the cable conductors, it decreases with increasing frequency and 
temperature. We can then expect that, if $T<T_{r}$, $\alpha(\nu,T)>1$ and, as can be seen in figure \ref{Gmtilde(nu)}, it will increase with frequency.

Let's now consider the difference between two determinations of $Z_{s}(\nu,T)$ with respect to  two different values of the temperature, say $T_{1}$ and $T_{2}$. Making use of equation \ref{Zs} , we can write
\begin{eqnarray}
\label{deltagamma0}
\Delta Z_{s}(\nu,T)\doteq Z_{1}(\nu,T_{1}) - Z_{2}(\nu,T_{2}) &=& Z_{0}\left(\frac{1+\Gamma_{0}(\nu,T_{1})}{1-\Gamma_{0}(\nu,T_{1})}-\frac{1+\Gamma_{0}(\nu,T_{2})}{1-\Gamma_{0}(\nu,T_{2})}\right)\nonumber \\
              &=& Z_{0}\left(\frac{1-\frac{\Gamma_{0}(\nu,T_{1})} {\Gamma_{0}(\nu,T_{2})}}{1+\frac{\Gamma_{0}(\nu,T_{1})}{\Gamma_{0}(\nu,T_{2})}-\frac{(1+\Gamma_{0}(\nu,T_{1}))(1+\Gamma_{0}(\nu,T_{2}))}{2\Gamma_{0}(\nu,T_{2})}}\right) \nonumber \\
           &\simeq& Z_0\frac{1-\Gamma_{0}(\nu,T_{1})/\Gamma_{0}(\nu,T_{2})}{1+\Gamma_{0}(\nu,T_{1})/\Gamma_{0}(\nu,T_{2})}
\end{eqnarray}
where the approximation involved in the last equivalence is possible if $\Gamma_0(\nu,T_2)\simeq -1$ (see below). Writing as usual $Z_{s}\doteq R_{s}+iX_{s}$ we have
\begin{equation}
    \Delta R_{s}(\nu,T)\simeq Z_{0}\frac{1-|\frac{\Gamma_{0}(\nu,T_{1})}{\Gamma_{0}(\nu,T_{2})}|^{2}}
{1+|\frac{\Gamma_{0}(\nu,T_{1})}{\Gamma_{0}(\nu,T_{2})}|^{2}+2Re[\frac{\Gamma_{0}(\nu,T_{1})}{\Gamma_{0}(\nu,T_{2})}]}
\label{Rsprima}
\end{equation}
\begin{equation}
    \Delta X_{s}(\nu,T)\simeq -2Z_{0}\frac{Im[\frac{\Gamma_{0}(\nu,T_{1})}{\Gamma_{0}(\nu,T_{2})}]}{1+|\frac{\Gamma_{0}(\nu,T_{1})}{\Gamma_{0}(\nu,T_{2})}|^{2}+2Re[\frac{\Gamma_{0}(\nu,T_{1})}{\Gamma_{0}(\nu,T_{2})}]}
\label{Xsprima}
\end{equation}
If the phase variation $\Delta\phi_{0}=\phi[\Gamma_{0}(T_{1})]-\phi[\Gamma_{0}(T_{2})]$ 
between  the two determinations of $\Gamma_{0}$ is small, we 
have $Re[\frac{\Gamma_{0}(\nu,T_{1})}{\Gamma_{0}(\nu,T_{2})}]\simeq 
\left|\frac{\Gamma_{0}(\nu,T_{1})}{\Gamma_{0}(\nu,T_{2})}\right|$ and equations \ref{Rsprima} and \ref{Xsprima} can be approximated as
\begin{equation}
    \Delta R_{s}(\nu,T)\simeq Z_{0}\frac{1-|\frac{\Gamma_{0}(\nu,T_{1})}{\Gamma_{0}(\nu,T_{2})}|}{1+|\frac{\Gamma_{0}(\nu,T_{1})}{\Gamma_{0}(\nu,T_{2})}|}
    \simeq Z_{0}\frac{1-\frac{|\alpha(\nu,T_{2})|}{|\alpha(\nu,T_{1})|}\frac{|\widetilde{\Gamma}_{m}(\nu,T_{1})|}
{|\widetilde{\Gamma}_{m}(\nu,T_{2})|}}{1+\frac{|\alpha(\nu,T_{2})|}{|\alpha(\nu,T_{1})|}\frac{|\widetilde{\Gamma}_{m}(\nu,T_{1})|}
    {|\widetilde{\Gamma}_{m}(\nu,T_{2})|}}
    \label{Rs0}
\end{equation}
\begin{equation}
    \Delta X_{s}(\nu,T)\simeq -2Z_{0}\frac{|\frac{\Gamma_{0}(\nu,T_{1})}{\Gamma_{0}(\nu,T_{2})}|}{(1+|\frac{\Gamma_{0}(\nu,T_{1})}{\Gamma_{0}(\nu,T_{2})}|) ^{2}}\Delta\phi_{0}
    \simeq -2Z_{0}\frac{\frac{|\alpha(\nu,T_{2})|}{|\alpha(\nu,T_{1})|}\frac{|\widetilde{\Gamma}_{m}(\nu,T_{1})|} {|\widetilde{\Gamma}_{m}(\nu,T_{2})|}}{(1+\frac{|\alpha(\nu,T_{2})|}{|\alpha(\nu,T_{1})|}\frac{|\widetilde{\Gamma}_{m}(\nu,T_{1})|}
    {|\widetilde{\Gamma}_{m}(\nu,T_{2})|})^{2}}(\Delta\phi_{m}-\Delta\phi_{\alpha})
    \label{Xs0}
\end{equation}
where we used the result stated in equation \ref{gamma0finale} into the last approximations. If we have $|\alpha(\nu,T_{1})|\simeq |\alpha(\nu,T_{2})|$, so that $\frac{|\Gamma_{0}(\nu,T_{1})|}{|\Gamma_{0}(\nu,T_{2})|}\simeq\frac{|\widetilde{\Gamma}_{m}(\nu,T_{1})|}{|\widetilde{\Gamma}_{m}(\nu,T_{2})|}$, equation \ref{Rs0} can finally be rewritten as
\begin{equation}
    \label{Rsm}
    \Delta R_{s}(\nu,T)\simeq Z_{0}\frac{1-\frac{|\widetilde{\Gamma}_{m}(\nu,T_{1})|}{|\widetilde{\Gamma}_{m}(\nu,T_{2})|}}{1+\frac{|\widetilde{\Gamma}_{m}(\nu,T_{1})|}{|\widetilde{\Gamma}_{m}(\nu,T_{2})|}}
\end{equation}
If we also have $\Delta\phi_{\alpha}\ll\Delta\phi_{0}$, so that $\Delta\phi_{0}\simeq\Delta\phi_{m}$, then equation \ref{Xs0} can be approximated with
\begin{equation}
    \label{Xsm}
    \Delta X_{s}(\nu,T)\simeq  -2Z_{0}\frac{\frac{|\widetilde{\Gamma}_{m}(\nu,T_{1})|}{|\widetilde{\Gamma}_{m}(\nu,T_{2})|}}{(1+\frac{|\widetilde{\Gamma}_{m}(\nu,T_{1})|}{|\widetilde{\Gamma}_{m}(\nu,T_{2})|}) ^{2}}\Delta\phi_{m}
\end{equation}

Formulae \ref{Rsm} and \ref{Xsm} are the main result of this paper. 
They relate the variation of impedance between two different temperatures to the measured quantities. This means that, if a reference temperature $T_{ref}$ exists for which 
$\Gamma_{0}\simeq -1$, we can extract from $\widetilde{\Gamma}_{m}$ the variations of  the curves $Z_{s}(\nu,T)$ with respect to $Z_{s}(\nu,T_{ref})$. 
Moreover, once all the $\Delta Z$ are known, if a temperature $T^{*}$ exist for which the curve $Z_{s}(\nu,T^{*})$ is known, we can get absolute values for $Z_{s}(\nu,T)$.
Finally, we have found two different formulae for the real and imaginary part of the impedance, so that we can handle them separately.

\subsection{Check of the validity of the approximations}

\begin{figure}
\centerline{\psfig{figure=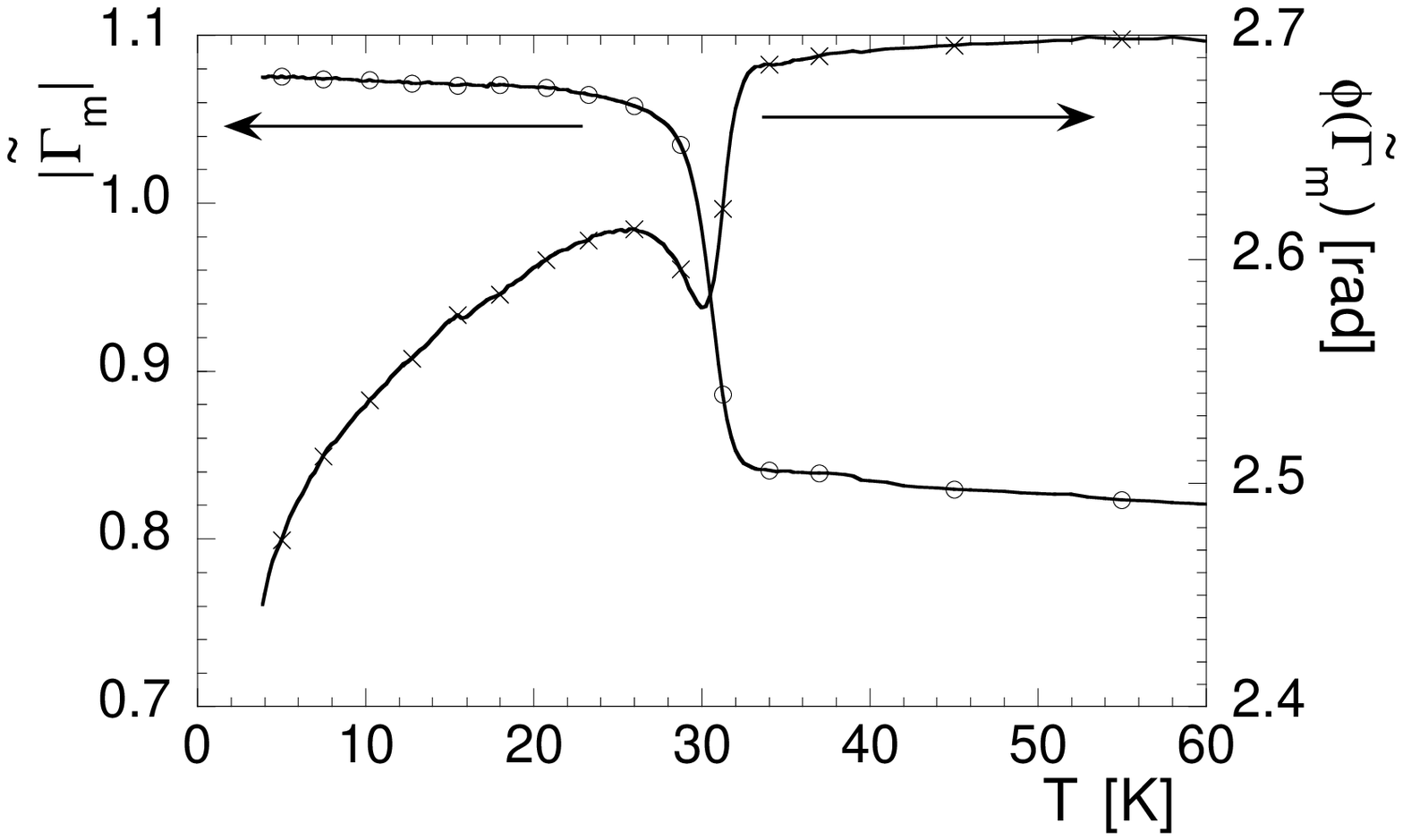,height=7.cm,width=11.5cm,clip=,angle=0.}}
\caption{Modulus (open circles, left axis) and phase (crosses, right axis) of $\widetilde{\Gamma}_{m}(\nu,T)$ at fixed frequency $\nu =10GHz$ as a function of temperature for $MgB_{2}$.}
\label{Gmtilde(T)}
\end{figure}

To employ equations \ref{Rsm} and \ref{Xsm} we must check that the approximations underlying them are all valid. For the sake of simplicity we analyze the specific case of $MgB_2$, being the procedure for $YBCO$ exactly the same.
First of all we will deal with the approximations on the temperature behaviour of $\alpha(\nu,T)$. In figure \ref{Gmtilde(T)} we show a typical curve at fixed frequency of the modulus and phase of $\widetilde{\Gamma}_{m}(T)$. In order to estimate the 
contribution to these curves due to $\alpha(\nu,T)$ we have to roughly guess the behaviour of $\Gamma_{0}(\nu,T)$, being from equation \ref{gamma0finale} 
$|\widetilde{\Gamma}_{m}(\nu,T)|=|\alpha(\nu,T)||\Gamma_{0}(\nu,T)|$ and $\phi(\widetilde{\Gamma}_{m})=\phi(\Gamma_{0})+\phi(\alpha)$. To do this, first we 
recall that $\Gamma_{0}(\nu,T)$ is linked to $Z_{s}(\nu,T)$ by relation \ref{Zs}, that can be inverted to
\begin{equation}
    \label{gammadizeta}
    \Gamma_{0}=\frac{Z_{s}-Z_{0}}{Z_{s}+Z_{0}}
\end{equation}
Above $T_{c}$ the resistivity is real, and so are the impedance and $\Gamma_0$. Being $Z_{s}<Z_{0}$ we have $\Gamma_{0}<0$ so that its phase is $\pi$.  Moreover above $T_{c}$ the resistivity is independent from frequency and we can compare microwave with DC measurements, finding that the resistivity is almost constant with temperature, so that $\Gamma_{0}(\nu,T\!\!>\!\!T_{c})$ is also nearly constant. Below the transition temperature the real part of the impedance attains a value much lower than the normal state one, so that we expect $R_{s}(\nu,T\!\!<\!\!T_{c})\simeq 0$, which, according to equation \ref{gammadizeta} implies 
$|\Gamma_{0}(\nu,T\!\!<\!\!T_{c})|\simeq 1$. As regarding the phase of $\Gamma_{0}(\nu,T\!\!<\!\!T_{c})$, using the two fluids model one can predict a peak near the transition temperature and a plateau sufficiently below it, whose value is very close to $\pi$.

We note in figure \ref{Gmtilde(T)} that $|\widetilde{\Gamma}_{m}(\nu,T)|$ is almost constant sufficiently above and below the transition temperature. Since $|\Gamma_{0}|$ is constant in these regions of temperature, we conclude that the small variations observed in $|\widetilde{\Gamma}_{m}(\nu,T)|$ are due to $|\alpha(\nu,T)|$. We can then state that $|\alpha(\nu,T)|$ does not appreciably vary across the whole curve so that we can neglect its variations with respect to those of $|\Gamma_{0}(\nu,T)|$ and apply equation \ref{Rsm}.

As regards the phase of $\alpha(T)$, it is evident from figure \ref{Gmtilde(T)} that the phase of $\widetilde{\Gamma}_{m}(T)$ shows a peak at the transition temperature that can be ascribed to $\phi(\Gamma_{0})$, but it is as much evident that $\phi(\widetilde{\Gamma}_{m})$ has not the same value above and below $T_{c}$. This means that the condition $\Delta\phi_{\alpha}\ll\Delta\phi_{0}$ doesn't hold and we can't make use of equation \ref{Xsm} to get sample reactance.
Nevertheless we can estimate from this graphic the maximum variation of $\phi(\Gamma_{0})$ in about 10 degrees, so that the approximation $|\Delta\phi_{0}|\simeq 0$ is well justified.

The last condition to be examined is the one employed in equation \ref{deltagamma0}: we need a reference temperature for which we are confident that $\Gamma_{0}(T_{ref})\simeq -1$. From the above discussion on $\Gamma_{0}$ and from figure \ref{Gmtilde(T)} we can state that $\Gamma_{0}\simeq -1$ for $T\leq 20K$. We choose $T_{ref}=4K$ for $MgB_{2}$ and, for analogous reasons, $T_{ref}=70K$ for $YBCO$.

\section{Results}

\begin{figure}
\centerline{\psfig{figure=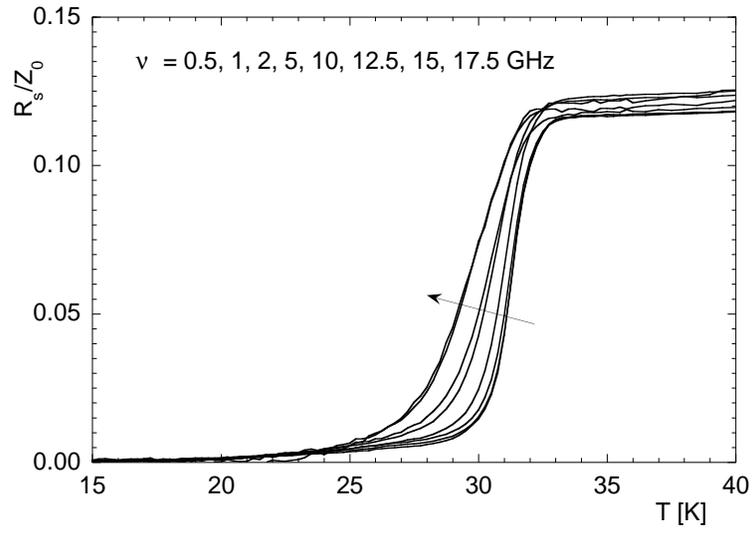,height=7.cm,width=10cm,clip=,angle=0.}}
\caption{Measured resistivities as a function of temperature, at several fixed frequencies. Note the widening of the transition, as the frequency is increased.}
\label{R1(T)_MgB2}
\end{figure}

\begin{figure}
\centerline{\psfig{figure=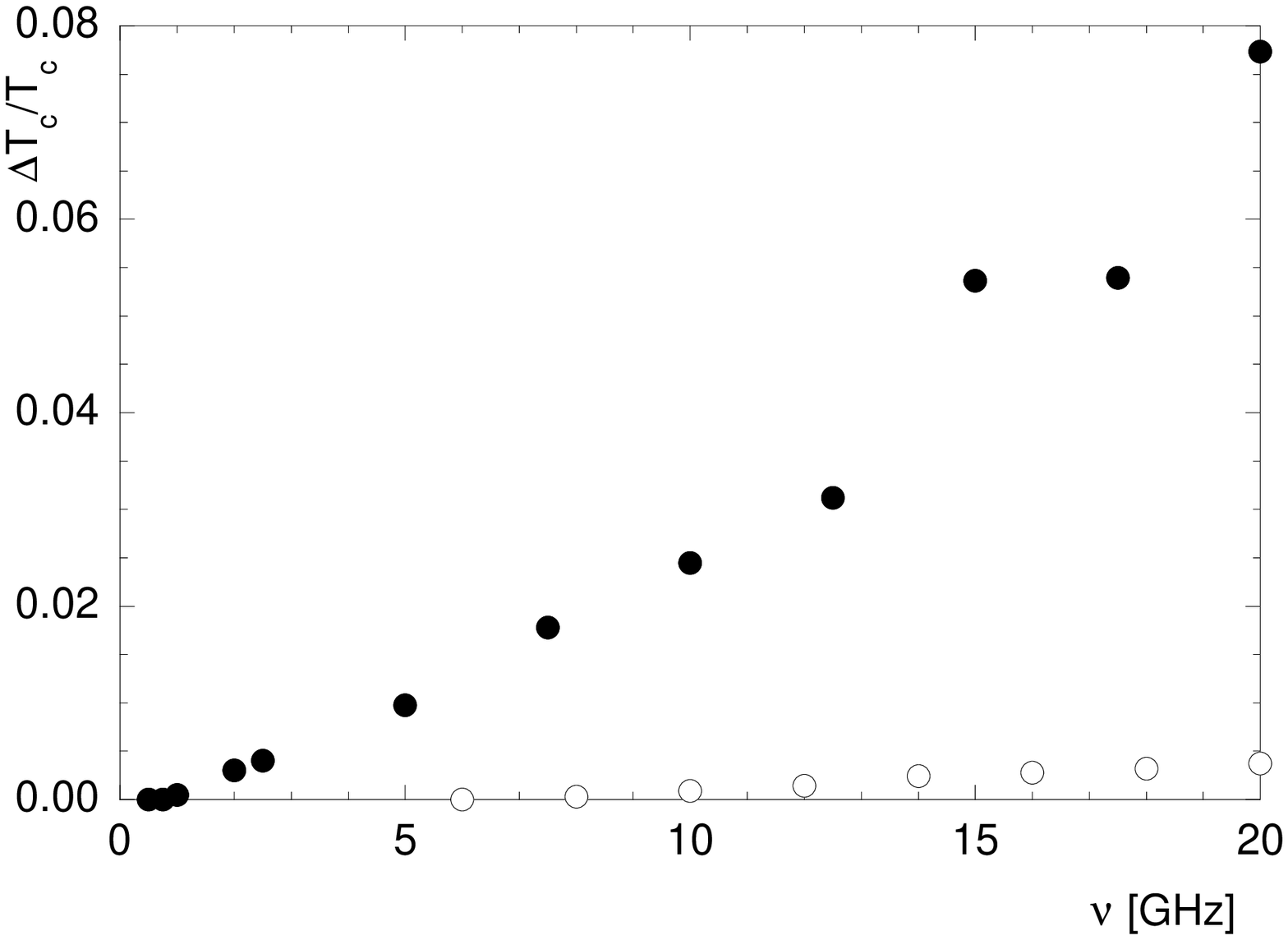,height=7.cm,width=9.cm,clip=,angle=0.}}
\caption{Widening of the transition as a function of frequency, for both $YBCO$ (open symbols) and $MgB_2$ (filled symbols) (see text for the definition of $\Delta T_c/T_c$)}
\label{Delta_Tc}
\end{figure}

The result of this procedure for the real part of $Z_s(T)$ is shown for the MgB$_2$ film in fig.\ref{R1(T)_MgB2}, where resistive transitions at different frequencies are plotted. The transition widens with increasing frequency, being nearly frequency independent up to $\simeq$ 2 GHz and then widening as the frequency is further increased. To have a more quantitative description of the widening, one can define $T_{mp}(\nu)$ as the temperature where the resistivity, at the given frequency $\nu$, reaches one half of the value in the normal state, and  $\Delta T_c (\nu) = T_{mp}(\nu) - T_{mp}(\nu_{min})$. The behaviour of $\Delta T_c(\nu)$, normalized to the value of $T_c$ at $\nu_{min} = 0.5$ GHz is reported in fig. \ref{Delta_Tc} (filled circles).

\begin{figure}
\centerline{\psfig{figure=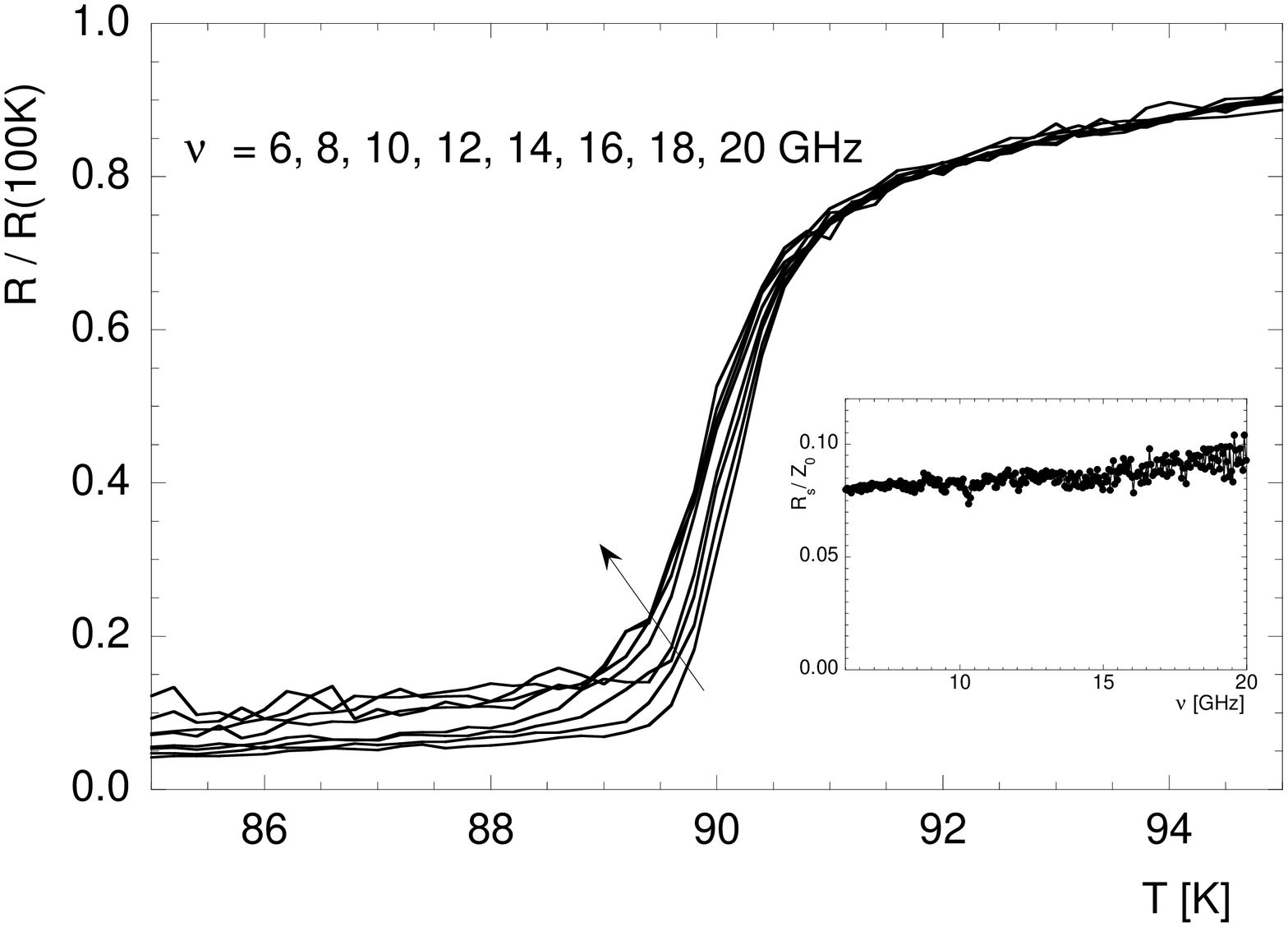,height=7.cm,width=9.cm,clip=,angle=0.}}
\caption{Measured resistivities as a function of temperature for the $YBCO$ film. The resistivities has been normalized to their value obtained at 100 K to reduce the effect of the variation of the cable parameters between 70 and 100 K. The behaviour of $R_s$ at 100 K is reported in the insert.}
\label{R1(T)_YBCO}
\end{figure}

The same analysis is presented in fig.\ref{R1(T)_YBCO}, for $YBCO$ where curves of $R_s/R_s(100K)$ are presented as a function of $T$ at various frequencies. The real part of the resistivity remains almost zero up to $\simeq 88\; K$, then grows at $T_c$ rapidly reaching the normal state value. As in the case of $MgB_2$, the transition widens as the frequency is increased. However, the widening is relatively small compared to $MgB_2$ and the upper part of the transition is almost unaffected by the variation of the frequency, being only the very last part of it severely dependent on frequency. The values of $\Delta T_c / T_c$ (with $\nu_{min} = 6$ GHz) are reported in fig. \ref{Delta_Tc}  for a quantitative comparison with the values obtained  for $MgB_2$.

\section{Conclusions}

We collected reflection coefficient measurements of  superconductive  thin films as a function of frequency through a Corbino disk technique. We developed a measurement technique by which we were able to obtain the resistivity from the reflection coefficients once that some conditions are verified. We extensively discuss these conditions and how we can conclude which of them are verified.
We found that not all of them are accomplished and we were only able to extract the real part of resistivity. The relative curves at fixed frequency are shown, together with a brief discussion of the data.

\newpage

\end{document}